\documentclass[11pt, a4paper]{article}
\usepackage[english]{babel}
\usepackage[utf8]{inputenc}
\usepackage[T1]{fontenc}
\usepackage{booktabs}
\usepackage{a4wide}
\usepackage{amsmath}
\usepackage{amsfonts}
\usepackage{url}
\usepackage[dvips]{graphicx}
\usepackage{calc}
\usepackage{subfigure}
\usepackage{multirow}
\usepackage{lettrine}
\usepackage{titlesec}
\usepackage{etoolbox}

\apptocmd{\sloppy}{\hbadness 10000\relax}{}{}
\newcommand{\half}[1]{\frac{#1}{2}}
\newcommand{\halb}{\half{1}}

\newcommand{\inv}[1]{\frac{1}{#1}}
\newcommand{\dd}{\:\text{d}}

\newcommand{\kzav}[1]{\left(#1\right)}
\newcommand{\hzav}[1]{\left[#1\right]}

\newcommand{\pd}[2]{\frac{\partial #1}{\partial #2}}

\newcommand{\rf}[1]{(\ref{#1})}

\providecommand{\keywords}[1]
{
  \small	
  \textbf{\textit{Keywords---}} #1
}

\title{Is The Internal Entropy of $F(R)$-Gravity Really An Entropy?}
\author{
	Bohuslav Matou\v{s} \\
	Observatory and Planetarium in Teplice\\
	Kopern\'{i}kova 3062, 415 01, Teplice\\
	Czech Republic\\
	}

\begin{document}
\maketitle
\abstract{This paper connects two methods for finding the functional of entropy in $F(R)$-Gravity: Padmanabhan's and Hammad's. The resulting approach is simple to follow and yields entropy functional, which can be separated into two parts. The part unknown in General Relativity is often called in the literature as an internal entropy and this paper points on incompatibility between the internal entropy found from the entropy functional and the one found using conventional approach. }


\section{Introduction}
$F(R)$-Gravity is a modified theory of gravity, where the scalar curvature in the Lagrangian is replaced with a general function $F(R)$ of the scalar curvature. This modification is simple because it does not introduce any other scalar as well as general as we have no conditions on the function $F(R)$, they are imposed in the moment when we compare predictions with observations \cite{DeFelice:2010aj, Nojiri:2010wj, Nojiri:2017ncd, Oikonomou:2022wuk}. Interesting is the fact that the $F(R)$-Gravity produces an analogy with thermodynamics similarly as General Relativity. Ever since the thermodynamics of $F(R)$-Gravity was studied \cite{Eling:2006aw}, the additional effects, arising from the higher order members of the function $F(R)$, were identified as disipative nature of gravity caused by a new entropy member --- internal entropy \cite{Chirco, Bamba:2009gq, Akbar:2006mq}. However, in a recent paper it was shown \cite{Matous:2021uqh}, that it might be more natural to see those effects as being caused by a pressure related to the function $F(R)$. Such ambiquity is due to similarity of terms $P\delta V$ and $T \delta S$ because they are almost equal in the effect. Nevertheless, even in the analogy between gravity and thermodynamics, the entropy is special and when into account works of Thanu Padmanabhan and Fay\c{c}al Hammad.

Thanu Padmanabhan held a specific view about nature of gravity. In his works \cite{Padmanabhan:2013nxa, Padmanabhan:2014jta, Padmanabhan:2015pza, Padmanabhan:2015zmr, Padmanabhan:2016bha}, he  and his team have shown, that if we assume that the space-time has a structure, the gravity is emergent as a consequence of thermodynamics of such structure. Hence, from his point of view, the gravity represents an effect of more fundamental fabric. In that case the behaviour of gravity is governed by the principle of maximizing the entropy functional. 

Fay\c{c}al Hammad has adopted a different approach \cite{Hammad:2010, Hammad:2014jya}. His starting point was also the theory of elastisticity. From it, he constructs the functional of entropy as well but with a different approach. Using the elasticity of space-time, he is able to obtain the same thermodynamic properties as it is usual in thermodynamics of gravity, e. g. the Bekenstein formula for entropy.

In this paper, it will be demonstrated how we can combine both approaches Padmanabhan's and Hammad's and how we can benefit from it. The functional of entropy will be constructed using combined approach. From the functional, we will identify the so-called internal part of entropy and try to follow the Hammad's approach and compare the results to previously obtained results of entropy \cite{Akbar:2006mq, Matous:2021uqh}. 

The paper is organized as follows, the second section introduces both approaches leading to entropy functional. The third section shows how we can combine those two approaches and how can we benefit from it. The fourth section uses the results of previous sections to identify the internal part of entropy and compare it to its expected form. This paper follows the East Coast convention with metric signature ($-,+,+,+$) and Latin indices running over 0..3 interval while the Greek ones over 1..3. The fundamental constants $c, G, \hbar, k_B$ are treated as equal to one.  

\section{The Tale of Two Methods}
\subsection{The Method of Padmanbhan}

In Padamanbhan's emergent gravity paradigm, the curvature of space-time represents an effect of a structure of space-time itself. Such structure exhibits thermodynamic behaviour and thus, if we are about to find a metric, it is given by maximizing the functional of entropy  \cite{Padmanabhan:2007en, Chakraborty:2015hna}
\begin{equation}\label{EG-entropyFunc}
S[\xi] = \int_\mathcal{V} \dd^D x \sqrt{-g} \kzav{ 
4 P_{ab}^{\;\;\,cd} \nabla_c \xi^a \nabla_d \xi^b - T_{ab} \xi^a \xi^b
}\:,           
\end{equation}
where $\mathcal{V}$ is the region of interest, $\xi^a$ are displacement vectors, and finally $P_{ab}^{\;\;\,cd}$ is the entropy tensor defined as
\begin{equation}
 P_{a}^{\;\,ijk} = \pd{L}{R^a_{ijk}} \:,
\end{equation}
where $L$ is Lagrangian. In General Relativity, it takes simple form of 
\begin{equation}\label{EG-entopyTensor}
  P_{a}^{\;\,ijk} = \inv{32\pi} \kzav{ 
  g^{ik} \delta^j_a - g^{ij} \delta^k_a
  }\:.
\end{equation}
Padmanabhan was studying Lancosz-Lovelock models of gravity, which are models where the Lagrangian can be expressed as \cite{Padmanabhan:2013xyr} $\mathcal{L} = Q_a^{bcd} R^a_{bcd}$ where $\nabla_c Q_a^{bcd} = 0 $. As our focus lies in $F(R)$-Gravity, we are limited with this approach, as only Lagrangian, which is both $F(R)$-Gravity and Lancosz-Lovelock, is the one of General Relativity. However, it has been shown \cite{Matous:2021uqh} that the entropy tensor plays a role also in $F(R)$-Gravity. That is why we can assume, that the results of Padmanabhan's approach can be beneficial for us.

In analyses of thermodynamics of $F(R)$-Gravity \cite{Akbar:2006mq, Matous:2021uqh}, we can see that the entropy consists of two parts. The part similar to the entropy of General Relativity, different only by factor $F'$, where prime denotes derivative with respect to $R$. And the whole new part, the so called internal entropy. Interestingly enough, the factor $F'$ is not just difference between entropies but also between entropy tensors, this is something we would expect in Lancosz-Lovelock gravity not $F(R)$-Gravity. Therefore, it seems natural that we would expect Padmanabhan-Entropy in the form of
\begin{equation}
S[\xi] = \int_\mathcal{V} \dd^4 x \sqrt{-g} \kzav{ 
4 P_{ab}^{\;\;\,cd} \nabla_c \xi^a \nabla_d \xi^b - T_{ab} \xi^a \xi^b + \mathcal{S}^{(int)}
}\:,           
\end{equation}
where we have added unspecified $\mathcal{S}^{(int)}$ responsible for internal entropy effects. Let us focus on the first part and use the entropy tensor of $F(R)$-Gravity
\begin{equation}
  P_{ab}^{\;\;\,cd} = \frac{F'}{32\pi} \kzav{ 
  \delta^c_a\delta^d_b  -\delta^d_a \delta^c_b 
  }\:,
\end{equation}
then we see that the entropy functional part of Padmanabhan-Entropy can be written as
\begin{equation} \label{Padmanabhan}
 4P_{ab}^{\;\;\,cd} \nabla_c \xi^a \nabla_d \xi^b = 
 \frac{F'}{8\pi} 
 \kzav{\delta^c_a\delta^d_b  -\delta^d_a \delta^c_b }
 \nabla_c \xi^a \nabla_d \xi^b =
 \frac{F'}{8\pi} 
 \kzav{g_{ki}g_{lm}  -g_{kl} g_{mi} }
 \nabla^i \xi^k \nabla^l \xi^m
 \:.
\end{equation}
This is all we can get so far. We have the entropy functional consisting of entropy tensor part and uspecified internal entropy part. Without knowledge of the second method we cannot move further.

\subsection{The Method of Hammad}
The other method was presented by Hammad \cite{Hammad:2010}, the foundation is the same -- the elasticity of space-time, but the approach is a bit different. Instead of focusing on Lagrangian, coefficients from classical theory of elasticity are introduced. Nevertheless, the entropy functional for General Relativity was found with this method and is identical to the one of Padmanabhan. The importance of this approach is that the theory of elasticity is further applied to achieve Bekenstein formula. This approach was further generalized to $F(\phi, R)$-Gravity \cite{Hammad:2014jya} where the functional of entropy was looked for in form of 
\begin{align}
S = \int_\mathcal{V} \dd^4 x \sqrt{-g} \left[ 
\kzav{ A g_{im} g_{kl} + B g_{il}g_{km} + C g_{ik} g_{lm}
} \nabla^i \xi^k  \nabla^l \xi^m  + \right. \nonumber\\
 \left. + \kzav{ D_{,i} g_{kl} + E_{,k} g_{il} + H_{,l} g_{ik}
}\xi^i  \nabla^k \xi^l 
+ \kzav{\lambda g_{ik} + T_{ik}} \xi^i \xi^k
\right]\:,\nonumber \\ \label{HammadEntropy}
\end{align}
where $A. B, C, D, E, H$ are scalar functions of $R$ and $\phi$. What is interesting, is that if we  choose to do only $F(R)$-Gravity, we would not find a unique Entropy functional, the addition of scalar field $\phi$ is required to find it.

\section{The Combined Method}
We have looked at two methods and we can see that \rf{Padmanabhan} reminds us strongly of \rf{HammadEntropy}. Now, we combine our knowledge obtained from Padmanbhan's approach into Hammad's one. Both approaches use different sign convention, thus we need to use negative entropy tensor. And therefore we know, that $B=0$, $A = -C = \frac{F'}{8\pi}$ and that $D, E,$ and $H$ are connected to the internal entropy $\mathcal{S}^{(int)}$. Then, we renormalize with $8\pi$ and construct the following entropy functional
\begin{align}
S = \int_\mathcal{V} \dd^4 x \frac{\sqrt{-g}}{8\pi} \left[ 
\kzav{F' g_{im} g_{kl} -  F' g_{ik} g_{lm}
} \nabla^i \xi^k  \nabla^l \xi^m  + \right. \nonumber\\
 \left. + \kzav{ D_{,i} g_{kl} + E_{,k} g_{il} + H_{,l} g_{ik}
}\xi^i  \nabla^k \xi^l 
+ \kzav{\lambda g_{ik} + 8\pi T_{ik}} \xi^i \xi^k
\right]\:.\nonumber \\
\end{align}
From now on, we will follow method of Hammad. But incorporating the entropy tensor gave us huge advantage because with it, we will be able to find all the coefficients without the necessity of another scalar field $\phi$. The variation of entropy is then
\begin{align}
 \delta S = \int_\mathcal{V} \dd^4 x \frac{\sqrt{-g}}{8\pi}
 \left\{ 
 \hzav{ 
 g_{lm} \kzav{D - H + 2F'}_{,k} - g_{lk} \kzav{D - H + 2F'}_{,m}
 } \nabla^l \xi^m
 + \right. \nonumber\\
 \left. + \kzav{ 
 2\lambda g_{ik} + 16\pi T_{ik} - \nabla_k D_{,i} - g_{ik} \Box E - \nabla_i H_{,k} - 2F' R_{ik}
 }\xi^i
 \right\} \delta \xi^k \:, \nonumber \\ 
\end{align}
where we have used the identity $\nabla_s\nabla_n \xi^a - \nabla_n \nabla_s \xi^a = R^a_{bsn} \xi^b$. As the entropy shall be extremal, we would like this variation to vanish. The first part looks quite easily, it vanishes when $D = H - 2F' + \beta$, where $\beta$ is a constant. Since in the definition of the entropy function we have a derivative of $D$ and not $D$ itself, we can just assume $\beta = 0$. The condition for vanishing of the entropy variation is then
\begin{equation}
 2\lambda g_{ik} + 16\pi T_{ik}  - g_{ik} \Box E - \nabla_k \nabla_i H - \nabla_i \nabla_k H + 2\nabla_k\nabla_i F' - 2F' R_{ik}
 = 0 \:,
\end{equation}
as a consistency check, we can see, that if we put General Relativity in ($F'=1, E=0, H=0$), the equation holds when $\lambda=\halb R$. Nevertheless, in the general $F(R)$-Gravity case, we need to do more steps. Since $H$ is scalar then $\nabla_k \nabla_i H = \nabla_i \nabla_k H$ and expressing the stress-energy tensor yields
\begin{equation}\label{se_tensor}
16\pi T_{ik} = 
- 2\lambda g_{ik} + g_{ik} \Box E + 2\nabla_k \nabla_i H - 2\nabla_k \nabla_i F' + 2F' R_{ik}
\:,
\end{equation}
then the energy conservation law can be expressed as
\begin{equation}
 0= 
- 2\nabla_{k} \lambda  + \nabla_{k} \Box E + 2\nabla_k\Box H + 2R_{ik} \nabla^i H - 2\nabla_k\Box F' + F' \nabla_k R
\:,
\end{equation}
where we have used the contracted Bianchi identity $\nabla_j R^j_k = \halb \nabla_k R$, and the identity $\Box\nabla_k a = R_{ik}\nabla^i a + \nabla_k \Box a$. From this, we see that thecondition divides into two parts as follows
\begin{equation}
 0= 
 \nabla_k \kzav{
 -2 \lambda  +  \Box E + 2 \Box H - 2\Box F' + F
 }
 + 2R_{ik} \nabla^i H
\:,
\end{equation}
to ensure the energy conservation, we need $H$ to vanish, then we obtain the following condition 
\begin{equation}
 0= 
 \nabla_k \hzav{
 -2 \lambda  +  \Box \kzav{ E - 2F' } + F
 }
\:,
\end{equation}
with solution $E = 2F' + \gamma$, $\lambda = \Lambda + \half{F}$, where $\gamma$ and $\Lambda$ are constants. But as was the case with $D$ and $\beta$, since the functional contains a derivative of $E$, we can put $\gamma$ to zero. Putting those parameters back into stress-energy tensor \rf{se_tensor} leads to Einstein equations for $F(R)$-Gravity with a cosmological constant $\Lambda$, which can be further assumed as zero 
\begin{equation}\label{eqMotion}
8\pi T_{ik} = 
 F' G_{ik} - \Lambda g_{ik} - \halb \kzav{F   -  F' R }g_{ik} - \nabla_k \nabla_i F' + g_{ik} \Box F' 
\:.
\end{equation}
The found functional of entropy is then 
\begin{align}\label{entropyFunc}
S = \int_\mathcal{V} \dd^4 x \frac{\sqrt{-g}}{8\pi} \left[ 
\kzav{F' g_{im} g_{kl} -  F' g_{ik} g_{lm}
} \nabla^i \xi^k  \nabla^l \xi^m  + \right. \nonumber\\
 \left. +2 \nabla^i\kzav{  F' g_{il} g_{km} -F' g_{ik} g_{lm}
}\xi^k  \nabla^l \xi^m 
+ \kzav{\half{F} g_{ik} + 8\pi T_{ik}} \xi^i \xi^k
\right]\:.\nonumber \\
\end{align}
The similarity of first two terms might give us a little hope that they both can be expressed using entropy tensor. Unfortunately, if one tries, she soon finds that it is not so. That means that the second term is not governed by the same principle as entropy tensor. The entropy may be expressed more explicitly as
\begin{align}\label{entropyFunc2}
S = \int_\mathcal{V} \dd^4 x \frac{\sqrt{-g}}{8\pi} \left[ 
F' \nabla_i \xi^k  \nabla_k \xi^i -  F' \nabla_i \xi^i  \nabla_k \xi^k
   + \right. \nonumber\\
 \left. +2 \xi_i \nabla_k \xi^i \nabla^k F' - 2 \xi_k \nabla_i \xi^i \nabla^k F'
+ \kzav{\half{F} g_{ik} + 8\pi T_{ik}} \xi^i \xi^k
\right]\:.\nonumber \\
\end{align}

\section{Internal Entropy}
In this section, we attempt to use the approach introduced by Hammad in \cite{Hammad:2010}, where it was shown how to obtain Bekenstein entropy from the entropy functional. Reproducing this approach we aim to inspect the internal part of entropy functional and compare the result with the one obtained by other approaches \cite{Akbar:2006mq, Matous:2021uqh}. To start, we need to separate the internal part of entropy from the entire functional of entropy. To do so, we first convert the first part in \rf{entropyFunc} into a divergence of a vector field, which yields
\begin{align}
S = \int_\mathcal{V} \dd^4 x \frac{\sqrt{-g}}{8\pi} \left[ 
\nabla_m\kzav{F' \xi^k \nabla_k \xi^m - F' \xi^m \nabla_k \xi^k
}  + \right. \nonumber\\
 \left. +2 \nabla^i\kzav{  F' g_{il} g_{km} - \halb F' g_{ik} g_{lm} - \halb F' g_{im} g_{kl}
}\xi^k  \nabla^l \xi^m 
+ \kzav{\half{F} g_{ik} + 8\pi T_{ik} - F'R_{ik}} \xi^i \xi^k
\right]\:,\nonumber \\
\end{align}
from this, the classical (external) entropy can be separated as
 \begin{align}
S^{(e)} = \int_\mathcal{V} \dd^4 x \frac{\sqrt{-g}}{8\pi} 
 \nabla_m  \kzav{F'\xi^k  \nabla_k \xi^m - F' \xi^m  \nabla_k \xi^k}    
\:,
\end{align}
where we see the integrand being the same as in GR case, except for the factor of $F'$. This is noticeable for two reasons. At first, because in the limit $F \rightarrow R$ it goes to GR expression. And at second, because the factor of $F'$ is in accord with our expectations based on the entropy tensor. The remaining part of the functional is the internal entropy. To simplify it, we will express $T_{ik}$  with the help of \rf{eqMotion}. Here, we shall note one important thing, the stress-energy tensor will cancel the expression in the bracket leaving only $g_{ik} \Box F' - \nabla_k \nabla_i F'$, which we have seen in \cite{Matous:2021uqh} that does not contribute to the thermodynamic equations at all, at least in the most simple cases. This is first disconnect between the two forms of internal entropy. Anyway, cancelling the stress-energy tensor yields the internal entropy in convenient form of another vector field divergence as
 \begin{align}\label{Si_final}
S^{(i)} = \int_\mathcal{V} \dd^4 x \frac{\sqrt{-g}}{8\pi} \nabla_i \left( 
  \xi_k \xi^k \nabla^i F' - \xi_k \xi^i \nabla^k F' 
\right)\:.
\end{align}

To be able to work further and process the integration, we need to express the displacement vector $\xi^i$. Hammad \cite{Hammad:2010} uses two conditions, firstly the tensor of local rotations is zero
\begin{equation}\label{localRot}
 \nabla^i \xi^j - \nabla^j \xi^i = 0 \:,
\end{equation}
and secondly there are no external forces in the generalized Hooke's law 
\begin{equation}\label{Hooke}
 -f^i = \mu \nabla^i \nabla_k \xi^k + \nu \kzav{\nabla_k \nabla^i \xi^k + \nabla_k \nabla^k \xi^i }
 \:,
\end{equation}
where $f^i$ are the external forces and $\mu$ and $\nu$ are Lamm\'{e} coefficients. The solution for displacements in \cite{Hammad:2010} is then found using Schwarzschild metric. However, in the framework of $F(R)$-Gravity, we cannot use that metric. For our purposes, it is much more suitable to use more general metric but still preserve the spherical symmetry as well as staticity and work with the following metric
\begin{equation} \label{metric}
 \dd s^2 = -a \dd t^2 + \inv{a} \dd r^2 + r^2 \dd \vartheta^2 + r^2 \sin^2 \vartheta \dd \phi^2\:,
\end{equation}
where $a$ is a function of $r$ which satisfies conditions $a(r_h) = 0$ and $a'(r_h) = 2 \kappa \neq 0$ for a parameter $r_h$ and a surface gravity $\kappa$. When we put this metric in the conditions \rf{localRot} and \rf{Hooke}, we soon find, that it is not possible to solve non-trivially both of them. This is not an issue of $F(R)$-Gravity, we might also take such metric in General Relativity and find the same obstacle. This forces us to look closely at those conditions. To do so, we will use the metric \rf{metric} and the Einstein equations of General Relativity. The vanishing tensor of local rotations is for us more fundamental than the Hooke's law. Therefore, we will sove it first and find a vector solving \rf{localRot} as
\begin{equation}
 \xi^0 = \frac{A}{a}\:,\quad
 \xi^1 = w(r)\:,\quad
 \xi^2 = \frac{B}{r^2}\:,\quad
 \xi^3 = 0\:,
\end{equation}
we will restrict ourselves on radial displacements, so we take $B=0$. Then we can proceed to calculations the forces of Hooke's law. From the equation \rf{Hooke}, we see that if we want external forces to vanish, it means that the part with coefficient $\mu$ as well as coefficient $\nu$ shall vanish separatelly, thus, we can solve for those parts separatelly. The $\mu$-part is non-zero only in radial coordinate and gives
\begin{equation}
 \partial^i \nabla_k \xi^k = a \partial_r \hzav{\inv{r^2}\kzav{wr^2}'} = 0\:,
\end{equation}
which holds, when $w(r) = \frac{C}{r^2} + Dr$. As we want displacements to vanish in infinity, we take $D=0$. Then we look at the $\nu$-part of Hooke's law
and find the non-zero components as
\begin{align}
 f^0 =  \frac{2A\nu}{a} \frac{2a'+a''r}{2r}\ =  \frac{2A\nu}{a} G^2_2\:, \quad
 f^1 = \frac{2C\nu}{r^2} G^2_2 \:, 
\end{align}
where $G^2_2$ is a component of the Einstein tensor. If we put the Einstein tensor to zero in General Relativity, then our metric \rf{metric} would change into Schwarzschild metric. However, in the $F(R)$-Gravity, we cannot follow this path and just assume the Einstein tensor to be zero, its non-zero components might get cancel by the higher curvature terms. Such higher curvature terms are acting as a pressure and energy of a scalar field and that is why it is fine, that we still have non-elastic forces present in the Hooke's law. So, for displacements in $F(R)$-Gravity we want no local rotation and vanishing $\mu$-part of generalized Hooke's Law. The $\nu$-part is non-zero and that is fine for us. The displacement vector has non-zero only temporal and radial part and they are the following
\begin{equation}
 \xi^0 = \frac{A}{a} \:, \quad \xi^1 = \frac{C}{r^2}  \:.
\end{equation}

Having found components of $\xi^i$, we can move further to the integration of entropy. Our goal is to compare the result of the integration to previously found result \cite{Matous:2021uqh} of the internal entropy to see whether they are in accord. The internal entropy is expected to take form of
\begin{equation}\label{deltaSi}
  \frac{\delta S^i}{\delta  r_h} = 
   \frac{A}{4} \frac{ F'_h R_h - F_h}{2\kappa} \:,
 \end{equation}
 where $A$ is the horizon surface, $F_h$ is value of $F(R)$ on the horizon, $F'_h$ is value of $F'(R)$ on the horizon and $R_h$ is value of scalar curvature on the horizon. The integral of \rf{Si_final} can be solved two ways, one can use that the divergence to transform it to three-dimensional integral
  \begin{align}
S^{(i)} = \int_{\delta\mathcal{V}} \dd^3 x \frac{\sqrt{n}}{8\pi} n_i \left( 
  \xi_k \xi^k \nabla^i F' - \xi_k \xi^i \nabla^k F' 
\right)\:,
\end{align}
but we will stick with the four-dimensional one where we can use that the divergence of a vector field $U^i$ is $\nabla_i U^i = \partial_i U^i + \frac{g_{,i}}{2g} U^i$, which rules out all parts of the vector potentional except for the radial one which is 
  \begin{align}
U^{1} = 
\xi_k \xi^k \nabla^1 F' - \xi_k \xi^1 \nabla^k F' =
  -A^2 \partial_r F' \:,
\end{align}
where we have used that $\nabla^k F' = a \partial_r F' \delta^k_1$ and $\xi_0\xi^0=-\frac{A^2}{a}$. The divergence is then
  \begin{align}
  \nabla_i U^{i} = 
-A^2 \partial^2_r F' - \frac{2A^2}{r} \partial_r F' = - \frac{A^2}{r^2} \partial_r(r^2 \partial_r F') \:.
\end{align}
The integral is then
  \begin{align}
S^{(i)} = \int  \dd r \dd t \dd \vartheta \dd \phi \frac{r^2\sin^2\vartheta}{8\pi}  \left[
  - \frac{A^2}{r^2} \partial_r(r^2 \partial_r F')
\right]\:,
\end{align}
which can be simplified into form of
  \begin{align}
S^{(i)} = - \frac{A^2}{2}  \int  
  r^2 \partial_r F' \dd t    
\:,
\end{align}
a look at the integrand finds that it depends on the second derivative of $F$. Variation with respect to $r_h$ might bring in even higher derivitaves. However, this is contradictory to the result of \rf{deltaSi}, where only the function $F$ and its first derivative was present. From this, we can say, that those two expression generally do not match. This of course does not cover some special cases of $F$. 

\section{Conclusions}
This article combined Padmanabhan's and Hammad's approaches of establishing an entropy functional. In the literature, we have seen that the entropy tensor plays some role in the entropy of $F(R)$-Gravity albeit it is not Lanczos-Lovelock gravity. By putting the entropy tensor into Hammad's form of entropy functional, we obtained simplified version of it and shown how to follow it to get the entropy. This entropy functional can be split into two parts. The classical entropy part can be expressed as a divergence of a vector field, it also fulfills our expectations based on entropy tensor. The part, which does not exist at all in General Relativity, shall be the internal part of entropy. The internal part can also be expressed as a divergence of another vector field. Although the method of comparing it to an expected result has its limits, we can see that those two results do not match. From it, we can see that there is certianly some internal entropy of $F(R)$-Gravity but it shall depend mostly on second derivatives of $F$. From this we might conclude that the internal entropy found in previous works is not an entropy at all and that its interpretation as a higher-curvature caused presure is more suitable.

\end{document}